\documentclass[preprint,12pt,sort&compress]{elsarticle}

\usepackage{amssymb}
\usepackage{amsthm}
\usepackage{amsmath}
\usepackage{mathrsfs}
\usepackage{graphicx}
\usepackage{epstopdf}
\usepackage{float}
\usepackage{caption}
\usepackage{subcaption}
\usepackage{bm}
\usepackage{bbm}
\usepackage{mathrsfs}
\usepackage{cleveref}
\usepackage{soul}
\usepackage{accents}
\usepackage{graphicx}
\usepackage{xcolor}
\usepackage{courier} 
\usepackage{listings} 
\usepackage[framed,numbered,autolinebreaks,useliterate]{mcode}

\journal{Advances in Engineering Software}

\makeatletter
\def\@author#1{\g@addto@macro\elsauthors{\normalsize%
    \def\baselinestretch{1}%
    \upshape\authorsep#1\unskip\textsuperscript{%
      \ifx\@fnmark\@empty\else\unskip\sep\@fnmark\let\sep=,\fi
      \ifx\@corref\@empty\else\unskip\sep\@corref\let\sep=,\fi
      }%
    \def\authorsep{\unskip,\space}%
    \global\let\@fnmark\@empty
    \global\let\@corref\@empty  
    \global\let\sep\@empty}%
    \@eadauthor={#1}
}
\makeatother

\begin{document}

\begin{frontmatter}



\title{Abaqus2Matlab: a suitable tool for finite element post-processing}


\author{George Papazafeiropoulos\fnref{NTUA}}

\author{Miguel Mu\~niz-Calvente\fnref{Uniovi}}

\author{Emilio Mart\'{\i}nez-Pa\~neda\corref{cor1}\fnref{DTU}}
\ead{mail@empaneda.com}

\address[NTUA]{Department of Structural Engineering, National Technical University of Athens. 15780 Zografou, Athens, Greece}

\address[Uniovi]{Department of Construction and Manufacturing Engineering, University of Oviedo, Gij\'{o}n 33203, Spain}

\address[DTU]{Department of Mechanical Engineering, Solid Mechanics, Technical University of Denmark, DK-2800 Kgs. Lyngby, Denmark}

\cortext[cor1]{Corresponding author.}

\begin{abstract}
A suitable piece of software is presented to connect Abaqus, a sophisticated finite element package, with Matlab, the most comprehensive program for mathematical analysis. This interface between these well-known codes not only benefits from the image processing and the integrated graph-plotting features of Matlab but also opens up new opportunities in results post-processing, statistical analysis and mathematical optimization, among many other possibilities. The software architecture and usage are appropriately described and two problems of particular engineering significance are addressed to demonstrate its capabilities. Firstly, the software is employed to assess cleavage fracture through a novel 3-parameter Weibull probabilistic framework. Then, its potential to create and train neural networks is used to identify damage parameters through a hybrid experimental-numerical scheme, and model crack propagation in structural materials by means of a cohesive zone approach. The source code, detailed documentation and a large number of tutorials can be freely downloaded from www.abaqus2matlab.com.
\end{abstract}


\end{frontmatter}


\section{Introduction}
\label{Introduction}

Partial Differential Equations (PDEs) govern the physics of most engineering systems. As analytical solutions are limited and generally restricted to idealized cases, the development of efficient and robust numerical methods marks a milestone in the solution of boundary value problems in structural mechanics, electromagnetism, heat transfer, mass diffusion and fluid dynamics, among many other disciplines. The Finite Element Method (FEM) has become the leading numerical technique for solving PDEs in the mechanical, civil, aeronautical and bioengineering industries. Among the wide range of packages available, Abaqus \cite{A16} is undoubtedly one of the most popular finite element tools for academics and practitioners.\\

However, practical applications often require considering non-linear conditions, where uncertainties hinder high fidelity numerical predictions. In such circumstances, the use of advanced analysis methodologies - such as inverse approaches, statistical tools or hybrid experimental-numerical techniques - has proven to compensate the lack of information, yielding results that are otherwise unobtainable. Matlab \cite{M16}, a multi-paradigm computing environment, is generally considered to be the most powerful software in this regard due to its advanced capabilities in statistics, machine learning, neural networks, curve fitting, model-based calibration and optimization. Yet, a connection between the two most used packages in, respectively, finite element modeling and mathematical analysis, is still lacking. To fill this gap, a novel software tool is here proposed: \emph{Abaqus2Matlab}, which allows to run Abaqus directly from Matlab and to post-process the results, providing a link between the two well-known packages in a non-intrusive and versatile manner. The present proposal enjoys the benefits of Matlab's user friendly and centralized environment, as opposed other powerful tools like Python, which require add-on libraries.\\

\emph{Abaqus2Matlab} is distributed as source code with the aim of facilitating research. Numerous codes have been made freely available through the years, positively impacting the computational mechanics community. For instance, Sigmund and co-workers presented an efficient topology optimization implementation \cite{S01,A11}, Bordas and collaborators \cite{B07,N08,M17} described an object-oriented programming library for the extended finite element method (X-FEM) and meshless methods, Giner \emph{et al.} \cite{G09} implemented the X-FEM in Abaqus through a user subroutine, Parks and Paulino \cite{PP12} described the numerical implementation of the PPR potential-based cohesive zone model, Nguyen \cite{N14} proposed an open source program to generate zero-thickness cohesive elements and Mart\'{\i}nez-Pa\~neda and Gallego \cite{MG15} provided a user subroutine to effectively define the material property variation of functionally graded materials in Abaqus. Other open-source software that has recently contributed to scientific progress includes FReET \cite{N14b}, a code to conduct statistical, sensitivity and reliability assessment; FraMePID-3PB \cite{L14}, a tool to identify fracture parameters in concrete through inverse analysis; NiHu \cite{FR14}, an open source C++ library for the boundary element method; ESFM \cite{HP14}, a general framework for meshless methods; NOSA-ITACA \cite{G15}, a finite element code for masonry structures; PCLab \cite{L15}, an object-oriented Monte Carlo/Finite Element software; and, $\mu$MECH \cite{S16}, an open source C/C++ library of analytical solutions to classical micromechanical problems.\\

The present manuscript is organized as follows. The software framework and architecture are explained in the following section. Then, Section \ref{Sec:Manual} provides usage instructions through simple examples. Section \ref{Sec:Applications} shows the capabilities of the toolbox by addressing two relevant engineering applications; namely, probabilistic analysis of cleavage fracture and inverse identification of damage parameters through neural networks. Finally, the work is summarized in Section \ref{Sec:Concluding remarks}. 

\section{Abaqus2Matlab}
\label{Sec:Abaqus2Matlab}

The main internal characteristics of \emph{Abaqus2Matlab} are described below. The structure of Abaqus results (*.fil) file is briefly described in the first place, as it is necessary to understand how the presented software stores Abaqus results. The reading procedure is then detailed and insight is given into the software architecture. 

\subsection{Creating and processing Abaqus' results (*.fil) file}

The results (*.fil) file can be used to transfer Abaqus analysis results to other packages. The aforementioned file can be written in binary or ASCII format, depending on the need for porting results between dissimilar operating systems. ASCII format is chosen in the present approach due to its versatility.

\subsubsection{Generation of Abaqus results (*.fil) file}

The Abaqus results file is obtained in ascii format by defining specific options in the input (*.inp) or restart (*.res) files. The results file generation procedure differs between Abaqus/Standard and Abaqus/Explicit, *FILE FORMAT, ASCII must be specified in the former and *FILE OUTPUT in the latter. The reader is referred to Abaqus documentation for more details.

\subsubsection{Output}

The following output types can be written to the results file: element, nodal, energy, modal, contact surface, element matrix, substructure matrix and cavity radiation factor. Nodes and elements are numbered globally in models that have been defined as an assembly of part instances. A map between user-defined numbers and internal numbers is printed to the data file (*.dat) if any output requested includes node and element numbers. Set and surface names that appear in the results file are given along with their corresponding assembly and part instance names, separated by underscores.

\subsubsection{Record format}

The results (*.fil) file is a sequential file that must be read up to the location of the desired data. All data items are converted into equivalent character strings and written in (logical) records. Each single line contains a series of 80 string characters, which may contain the full record or part of it. In the latter case, after completely filling the first line, the record string continues at subsequent lines. The beginning of each record is indicated by an asterisk (*) and the data items are arranged immediately behind each other within each record. Each record has the format shown in Table \ref{tab:FormatRecord}.

\begin{table}[H]
\caption{Format of a record written in an Abaqus results file}
\centering
\begin{tabular}{c c c} 
\hline
 Location & Length & Description\\
 \hline
 1 & 1 & Record Length (L)\\
 2 & 1 & Record type key\\
 3 & (L-2) & Attributes \\
 \hline
\end{tabular}
\label{tab:FormatRecord}
\end{table}

The location number denotes the position in the record where a series of consecutive data items are written. The number of data items in each series is denoted by the length number. The first data item is an integer denoting the number of data items in the record. The second one defines the record type key, an indicator denoting the type of data. And finally the attributes are contained in a series of L-2 data items, at the 3$^{rd}$ position of a record. 

\subsubsection{Data item format}

Integer numbers are denoted by the character I, followed by a two digit integer which shows the number of the digits of the integer with the value of the integer following. On the other hand, floating point numbers begin with the character D, followed by the number in the format E22.15 or D22.15, depending on the precision. And character strings begin with the character A, followed by eight characters. If the length of a character string is less than 8, then the trailing positions are filled with blank spaces. If the length of a character string is larger than 8, then the character string is written in consecutive character strings, eight characters at a time.

\subsection{Reading Abaqus results files with Abaqus2Matlab}

A function named \mcode{Fil2str} is defined in Matlab to read the Abaqus results (*.fil) file by considering the data as a string and concatenating lines horizontally, as shown in listing \ref{lst:Fil2str}.
\newpage

\begin{footnotesize}
\begin{lstlisting}[caption = {Function Fil2str.m to read Abaqus results (*.fil) file},label={lst:Fil2str}]
function Rec = Fil2str(ResultsFileName)
% Open the results file for reading
fileID = fopen(ResultsFileName,'r');
% Read data as a string and assign them to a cell array
% Concatenate each line without specifying delimiter, white 
% space or end of line characters
try
C = textscan (fileID, '%s', 'CollectOutput', '1', ...
        'delimiter','','whitespace','','endofline','');
catch
C = textscan (fileID, '%s', 'CollectOutput', 1, ...
        'delimiter','','whitespace','','endofline','');
end
% Close the results file
fclose(fileID);
% Assign A
A = C{1}{1};
% Remove newline characters
A1 = strrep(A,sprintf('\n'),'');
% Remove carriage return characters
Rec = strrep(A1,sprintf('\r'),''); 
\end{lstlisting} 
\end{footnotesize}\vspace{10pt}

The function is programmed so as to allow compatibility between different MATLAB versions. The information from the results file is stored in a cell array \mcode{C} containing a single line string. That single line string subsequently enters an \emph{ad hoc} function that depends on the results that the user wishes to post-process. Thus, more than 50 different functions are already available in \emph{Abaqus2Matlab}, covering the vast majority of results types that can be obtained in Abaqus; new record functions can be easily generated from the existing template. An appropriate naming convention is adopted, where each function is defined by the word \mcode{Rec} followed by the record key of the particular type of results. Record keys for each specific set of results can be found in Abaqus documentation. For example, nodal coordinates (record key 1901) are obtained through function \mcode{Rec1901.m}, whose code is shown in listing \ref{lst:Rec}.\\

The programming of such functions follows a similar structure. First, the record length is obtained for preallocation purposes, using Matlab's intrinsic function \mcode{strfind} to find the positions of the records in \mcode{Rec}. For each case, the first 8 characters in front of each position are stored in \mcode{Rec2}. Afterwards, the record length is identified and stored in the column vector \mcode{NW} by first converting from string to double format using Matlab's built-in function \mcode{str2num}. Subsequently, the elements of \mcode{ind} (i.e. position of the second data item of records giving node definition results) are scanned and for each element the number of digits of the node number is determined first, then the node number, and finally the nodal coordinates, by the insertion of a \mcode{for} loop within each record definition. Finally, the node numbers and the node coordinates are concatenated horizontally to form the output array. Hence, the \mcode{Rec1901} function takes as input a one-row string containing the ASCII code of the ABAQUS results (*.fil) file and provides as output a matrix with as many rows as nodes in the model and with the node number in the first column and the nodal coordinates in the subsequent columns.
 
\newpage
\begin{footnotesize}
\begin{lstlisting}[caption = {Example of a specific record function},label={lst:Rec}]
function out = Rec1901(Rec)
ind = strfind(Rec,'I 41901'); % record key for node output
if isempty(ind)
    out=[];return;
end
nextpos=numel('I 41901')+1;
% Initialize
NodeNum=zeros(numel(ind),1);
NW=zeros(numel(ind),1);
for i=1:numel(ind)
    % find the record length (NW)
    Rec2=Rec(ind(i)-7:ind(i));
    indNW=strfind(Rec2,'*'); % record starts with *
    % ensure record existence and appropriate type key location
    if isempty(indNW) || indNW>3
        ind(i)=NaN;continue;
    end
    % number of digits of the record length
    ind1=indNW+2; ind2=indNW+3;
    a1=str2num(Rec2(ind1:ind2));
    % Record length (NW)
    ind1=ind1+2; ind2=ind2+a1;
    NW(i)=str2num(Rec2(ind1:ind2));
end
NodeCoords=zeros(numel(ind),max(NW)-4);
for i=1:numel(ind)
    % number of digits of the node number
    ind1=ind(i)+nextpos; ind2=ind(i)+nextpos+1;
    a1=str2num(Rec(ind1:ind2));
    % Node number
    ind1=ind1+2; ind2=ind2+a1;
    NodeNum(i)=str2num(Rec(ind1:ind2));
    % Node coordinates
    for j=1:NW(i)-4
        % node coordinate
        ind1=ind2+2;
        ind2=ind2+23;
        NodeCoords(i,j)=str2num(Rec(ind1:ind2));
    end
end
% Assemply of matrices for output
out=[NodeNum NodeCoords];
end
\end{lstlisting}
\end{footnotesize}

\section{Usage instructions}
\label{Sec:Manual}

A brief description of the source code assembly is given first to ease the understanding of the software and allow for personalized developments. The most relevant operations are then detailed and finally a simple optimization example is described to show the use of \emph{Abaqus2Matlab}. Comprehensive documentation can be found in www.abaqus2matlab.com.

\subsection{Organization of the source code}
\label{Sec:Sourcecode}

Source code files are organized through the following folders:
\begin{itemize}
\item OutputAnalysis folder, which contains the functions required to post-process analysis type results (e.g. node definitions, element connectivity, eigenfrequencies and eigenvalues, etc.).

\item OutputNodes folder, which contains the functions required to post-process nodal type results (e.g. node displacements, concentrated forces, nodal temperatures, etc.).

\item OutputElements folder, which contains the functions required to post-process element type results, i.e. results at integration points or centroids (e.g. stresses, strains, section forces and moments, etc.).

\item Verification folder, which contains numerous Matlab scripts to verify the \mcode{Rec} functions corresponding to each result type.

\item AbaqusInputFiles folder, which contains the input files that are run by Abaqus to verify each of the results functions. 

\item Html and Help folders, which contain all the documentation files of \emph{Abaqus2Matlab}. 

\end{itemize}

\subsection{Main usage instructions}
\label{Sec:Instructions}

Firstly, the \mcode{Documentation.m} file must be compiled in order to add all \emph{Abaqus2Matlab} specific functions to Matlab's libraries. Emphasis has been placed in the development of user-friendly and intuitive software. As a consequence, only three steps are required; namely, (i) generate Abaqus' input file, (ii) run the finite element analysis and (iii) extract the results required. Thus, one should first indicate in Abaqus' input file the specific data that should be stored in the results file (*.fil); e.g. nodal displacements,\\

\begin{lstlisting}[caption = {Commands required to store the specific information in the *.fil file},label={lst:AbaqusInput},commentstyle=\small\textit]
*FILE FORMAT, ASCII
*NODE FILE
U
\end{lstlisting}\vspace{10pt}

\noindent the code described in Listing \ref{lst:AbaqusInput} should be included at the end the step definition (i.e., at the end of the document if it is a one-step analysis), just before the command *END STEP.\\

The remaining steps are performed entirely inside \emph{Abaqus2Matlab} framework within Matlab. A template script is shown in listing \ref{lst:Template}. First, the finite element job is run for a specific Abaqus input file. Both the script file and Abaqus' input file must be stored in the same folder (i.e., the working directory). The information in the results file (*.fil) is then read and classified according to the desired output data through \mcode{Fil2str} and - for the displacement field - \mcode{Rec101}, respectively. 

\begin{footnotesize}
\begin{lstlisting}[caption = {Example of a specific \emph{Abaqus2Matlab} script},label={lst:Template}]
%% Template script Abaqus2Matlab
% Run the corresponding input file 101.inp with Abaqus
Inp_file='101';
system(['abaqus job=' Inp_file]);
% Pause Matlab execution to create the lck file
pause(10)
% If the lck file exists then halt Matlab execution
while exist([Inp_file '.lck'],'file')==2
pause(0.1)
end
%% Postprocess Abaqus results file with Matlab
% Assign all lines of the fil file in an one-row string
Rec = Fil2str( [Inp_file '.fil']);
% Obtain the desired output data
Displacements= Rec101(Rec);
\end{lstlisting}
\end{footnotesize}\vspace{10pt}

Template scripts for the most often used types of Abaqus results have been developed and can be found in the Verification folder. 

\subsection{Simple example: truss optimization problem}
\label{Sec:Example}

\begin{figure}[H]
\centering
\includegraphics[scale=0.9]{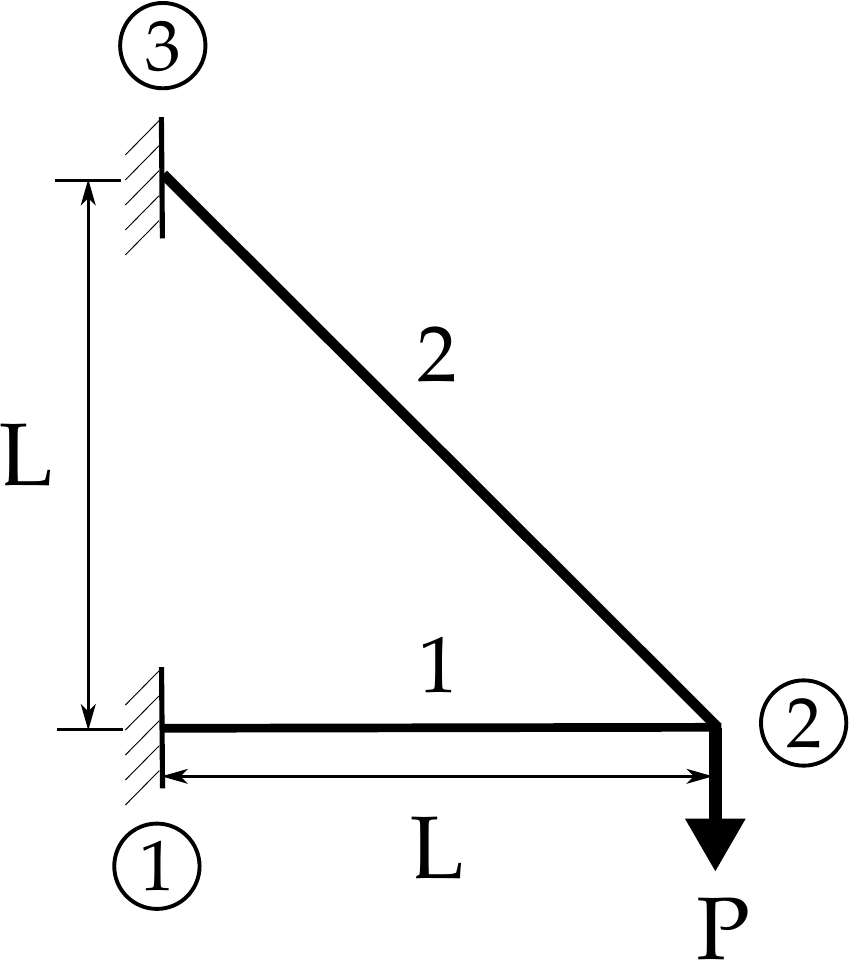}
\caption{Optimization example: a 2-bar plane truss, including element and node (circles) numbering.}
\label{fig:Optimization}
\end{figure}

A simple optimization example is shown to display \emph{Abaqus2Matlab} functioning. The weight of the 2-bar plane truss shown in Fig. \ref{fig:Optimization} will be optimized by minimizing the members' cross-sectional area. The truss is characterized by the following quantities: Young's modulus $E=68.948$ GPa, density $\rho=2767.99$ kg/m$^3$, bar length $L=9.144$ m and applied concentrated forces $P=444.974$ kN. Material costs can be lowered by reducing the weight, which implies - for a constant density and bar length - minimizing the sum of the cross-sectional areas, the design variables under consideration. Constraints are imposed on the displacements, where their maximum value should be limited to $d_{max}=0.0508$ m in both directions, and stresses, where their magnitudes should be lower than $\sigma_{max}=172.369$ MPa in both tension and compression (absolute value). The design variables are the cross section area of each member in the interval [0.00365, 0.02258] m$^2$, with the lower bound being a consequence of the upper limit imposed on the axial stress.

The main script employed to solve the problem is shown in listing \ref{lst:MainOpt}. The number of elements is first specified and an initial guess for the 2 cross-sectional areas assigned. Before calling the main optimization function, design variable limits are defined and tolerances provided; the latter include an upper bound on the variation of the objective function (i.e., truss weight) during a step and an upper bound on the magnitude of the constraint functions.

\begin{footnotesize}
\begin{lstlisting}[caption = {Main code for the optimization example},label={lst:MainOpt}]
% Specify the number of elements of the truss
NumElements=2;
% Make a starting guess for the solution.
x0 = [0.0037; 0.0049];
% Set the lower and upper limits
AreaMin=0.003650822800775; % P*sqrt(2)/maxstress
AreaMax=0.0225806;
lb=AreaMin*ones(1,NumElements);
ub=AreaMax*ones(1,NumElements);
% Set FunctionTolerance and StepTolerance
options=optimset('fmincon');
options.Display='iter-detailed';
options.TolFun=1e-3;
options.TolCon=1e-3;
% Perform constrained optimization of the truss
[X,fval,exitflag,output,lambda]=fmincon(@TrussObjfun,x0,...
    [],[],[],[],lb,ub,'TrussConfun',options)
\end{lstlisting}
\end{footnotesize}\vspace{10pt}

\mcode{TrussObjfun} (listing \ref{lst:TrussObjfun}) is a simple function that provides as output the weight of the truss for given values of the design variables. \mcode{TrussConfun.m}, shown in listing \ref{lst:TrussConfun}, is employed to construct the Abaqus input file and subsequently perform the calculations. The input file is created through \mcode{TrussInpFileConstr.m} on every constraint evaluation, changing the data lines corresponding to the cross section area. Next, postprocessing of the aforementioned results takes place, which concludes in the formation of the inequality and equality vectors required as an output of the constraint function in Matlab (\mcode{c} and \mcode{ceq}, respectively). All the required information is then available for Matlab's function \mcode{fmincon} to perform the optimization analysis.

\begin{footnotesize}
\begin{lstlisting}[caption = {TrussObjfun function},label={lst:TrussObjfun}]
function f = TrussObjfun(x)
u=9.144; % Horizontal length
f = 9.81*2767.990471*x'*u*[1;sqrt(2)]; % total weight
end
\end{lstlisting}
\end{footnotesize}\vspace{10pt}

Results show that design variables at the local minimum equal [0.00365, 0.00482] m$^2$, while the minimum truss weight renders 2.5987 kN. For the particular example under consideration, this has been achieved after 6 iterations and 21 objective function evaluations. Optimization methodologies, like the one outlined here, can be used at the concept stage of the design process to achieve a proposal that best fits performance and manufacturability requirements. Such tools have proven to substantially reduce design development costs by avoiding expensive and time consuming design iterations.\\

\begin{footnotesize}
\begin{lstlisting}[caption = {Constraint function TrussConfun.m},label={lst:TrussConfun}]
function [c,ceq] = TrussConfun(x)
% Set the displacement limits of the 2-bar truss
Dmaxhor=0.0508;Dmaxver=0.0508;
% Construct the Abaqus input file TrussABAQUS.inp
TrussInpFileConstr(x)
% Run the input file TrussABAQUS.inp with Abaqus
!abaqus job=TrussABAQUS
% Pause Matlab execution to create the TrussABAQUS.lck file
pause(10)
while exist('TrussABAQUS.lck','file')==2
    pause(0.1)
end
% Assign all lines of the TrussABAQUS.fil file in an one-row string
Rec = Fil2str('TrussABAQUS.fil');
% Obtain the nodal displacements
out2 = Rec101(Rec);
NodalDisplacements=out2(:,2:3);
% Delete the files of last Abaqus run to avoid rewriting them
delete('TrussABAQUS.fil');delete('TrussABAQUS.prt');
delete('TrussABAQUS.com');delete('TrussABAQUS.sim');
% Calculate the maximum nodal displacements
maxNodDisplX1=max(abs(NodalDisplacements(:,1)));
maxNodDisplY1=max(abs(NodalDisplacements(:,2)));
% Assemble the constraints
c = [maxNodDisplY1-Dmaxver;
    maxNodDisplX1-Dmaxhor];
ceq = [];
end
\end{lstlisting}
\end{footnotesize}\vspace{10pt} 

Results can be obtained in a few minutes and all the necessary files can be freely downloaded from \emph{Abaqus2Matlab} website.

\section{Applications}
\label{Sec:Applications}

The range of applications of \emph{Abaqus2Matlab} is enormous, as it provides a non-intrusive connection between a sophisticated finite element package and the most comprehensive mathematical analysis tool. For demonstration purposes, two problems of particular interest from the scientific and engineering perspective will be addressed here. On the one hand, the toolbox proposed is used to estimate cleavage fracture in metals, where a probabilistic approach is needed due to the statistical nature of the micromechanisms involved. On the other hand, an advanced inverse-optimization methodology is employed to obtain the parameters governing the traction-separation law that describes deformation and fracture.

\subsection{Cleavage fracture}

Cleavage fracture originates from microcracks that nucleate from defects (carbides, cracks arrested at grain boundaries, etc.). The location of these defects is statistical by nature and hence modeling efforts rely on probabilistic analysis. Since the seminal work by Beremin \cite{B83}, cleavage fracture toughness estimations are based on Weibull statistics and the weakest link model, where the probability of failure equals the probability of sampling (at least) one critical fracture-triggering particle. Grounded on this approach, we propose a novel probabilistic framework that takes advantage of the advanced statistical tools of MATLAB to estimate all Weibull-related parameters without any \emph{a priori} assumptions.\\

For a given Weibull stress $\sigma_w$ and a threshold stress for crack growth $\sigma_{th}$, the cumulative probability of failure $P_f$, in terms of the modulus $m$ and scaling parameter $\sigma_u$ is given by,
\begin{equation}\label{Eq:Pf}
P_f=1-\exp \left[- \left( \frac{\sigma_w-\sigma_{th}}{\sigma_u} \right)^m \right]
\end{equation}

\noindent where the Weibull stress can be defined as,
\begin{equation}\label{Eq:Sw}
\sigma_w = \sigma_{th} + \left[ \sum_{i=1}^{n_e} \left( \sigma_1^i - \sigma_{th} \right)^m \left(V_i / V_0 \right) \right]^{(1/m)}
\end{equation}

Here $V_0$ is a reference volume, $V_i$ is the volume of the i\emph{th} material unit (finite element) in the fracture process zone experiencing a maximum principal stress $\sigma_1^i$ and $n_e$ is the number of finite elements/material units in the fracture process zone. The parameter $\sigma_{th}$ is needed due to the fact that cracks do not propagate below a certain threshold energy value. However, the concurrent estimation of the threshold, modulus and shape parameters remains a complicated task; a common approach lies in assuming a value for $\sigma_{th}$ and estimating $m$ and $\sigma_u$ from a set of experiments. Here, all three parameters ($\sigma_{th}$, $m$ and $\sigma_u$) will be obtained by means of a novel iterative procedure involving least squares estimates of the cumulative distribution functions.\\

The capabilities of \emph{Abaqus2Matlab} to model cleavage fracture will be benchmarked with an extensive experimental data set developed within the Euro toughness project \cite{HH02}. As in the experiments, a quenched and tempered pressure vessel steel DIN 22NiMoCr37 steel will be investigated; only tests where significant ductile crack growth is not observed will be considered and the reference experimental data will be that obtained at -40$^{\circ}$C with a compact tension specimen of size 1T. Comprehensive details of the material tensile properties, specimen size and failure loads are provided in the original experimental article \cite{HH02} and will not be reproduced here for the sake of brevity.\\

\begin{figure}[H]
\centering
\includegraphics[scale=0.8]{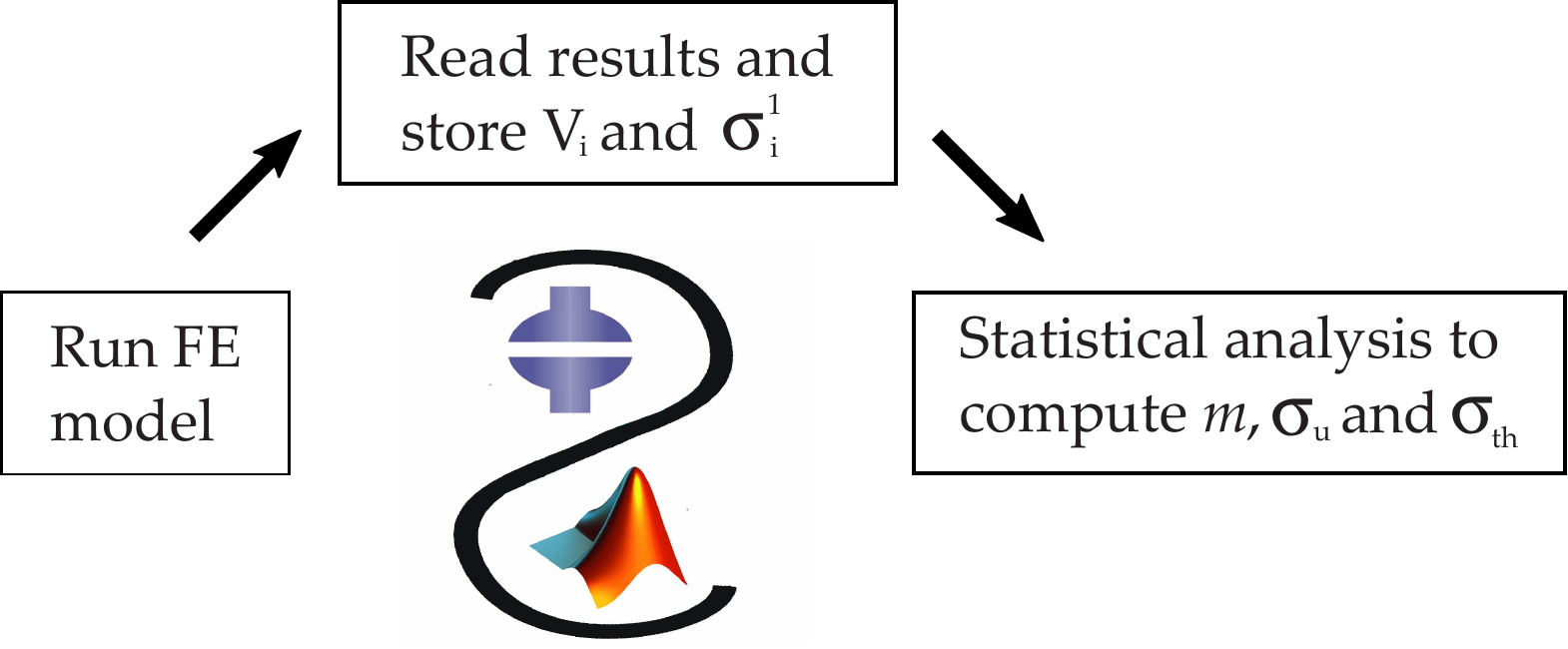}
\caption{Schematic overview of the use of \emph{Abaqus2Matlab} to estimate the probability of cleavage failure.}
\label{fig:Weibull1}
\end{figure}

The algorithm methodology is described in Fig. \ref{fig:Weibull1}. First, the finite element results are computed by running the corresponding Abaqus job \emph{inside} the proposed toolbox. A finite element mesh of approximately 2000 quadratic plane strain quadrilateral elements with reduced integration is employed, with the elements being progressively smaller as the crack tip is approached. The values of the volume element $V_i$ and the maximum principal stress $\sigma_1^i$ are read and stored for each finite element and load level of interest. The latter is characterized through the $J$-integral and the pin displacement, that are also read in the \emph{Abaqus2Matlab} environment. The statistical analysis is then conducted. The probability of failure for each failure load reported experimentally is first computed through,
\begin{equation}
P_f=\frac{j-0.3}{n_j + 0.4}
\end{equation}

\noindent where $n_j$ denotes the number of experiments and $j$ the rank number. Afterwards, an iterative procedure is conducted to simultaneously estimate $\sigma_{th}$, $m$ and $\sigma_u$. In each iteration the Weibull stress is computed from the values of $m$ and $\sigma_{th}$ from the previous iteration and subsequently inserted in Eq. (\ref{Eq:Pf}) to compute the values of $\sigma_u$, $m$ and $\sigma_{th}$ in the current iteration by fitting a univariate distribution through the least squares method. Convergence is achieved when the relative norm of the change in the solution is below an appropriate tolerance value. Therefore, taking advantage of Matlab capabilities, Weibull parameters are calculated by finding the distribution whose cumulative function best approximates the empirical cumulative distribution function of the experimental data. 

\begin{figure}[H]
\centering
\includegraphics[scale=0.9]{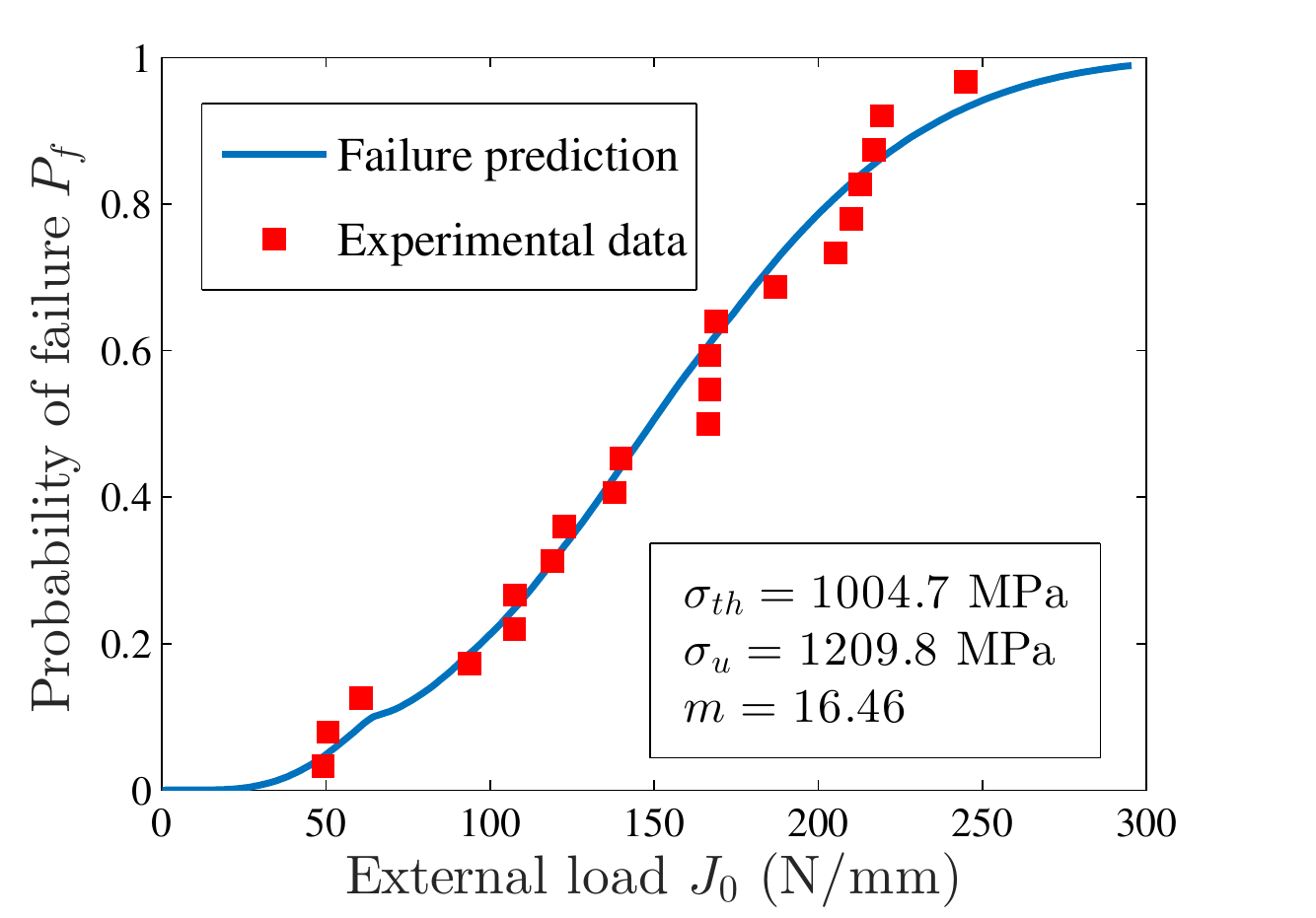}
\caption{Failure probability as a function of the external load. The figure includes the experimental data for 22NiMoCr37 steel (EuroData project \cite{HH02}) and the predictons from the present statistical model for the values of $\sigma_{th}$, $\sigma_u$ and $m$ displayed.}
\label{fig:WeibullLoad}
\end{figure}

The results obtained for the particular case considered (Euro toughness data set, T1, -40$^{\circ}$C) are displayed in Fig. \ref{fig:WeibullLoad}. The figure shows the probability of failure versus the external load from the experimental study and the
current statistical model. The calibrated Weibull stress parameters are also embedded in the figure. As it can be observed, a good agreement is attained between the failure probability estimated from Eq. (\ref{Eq:Pf}) and the experimental results. 

Results indicate that, for the particular case under consideration, a 50\% probability of failure will be attained for an external load of approximately $J_0=150$ N/mm, while the 5\% and 95\% probability bonds are attained at $J_0=60$ N/mm and $J_0=250$ N/mm, respectively. Weibull-parameters estimation reveals that stresses lower than $\sigma_{th}=1004.7$ MPa are innocuous and that a failure probability of 64\% in a unit element is attained at a stress level of $\sigma_u + \sigma_{th} = 2214.5$ MPa.\\

\begin{figure}[H]
\centering
\includegraphics[scale=0.9]{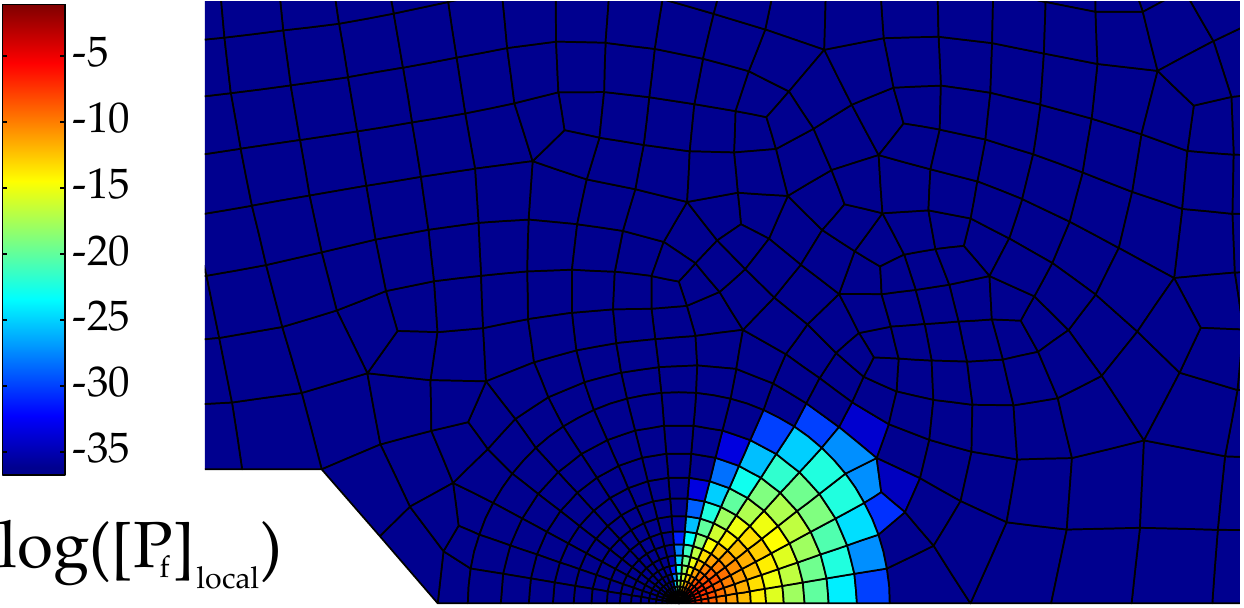}
\caption{Hazard map. The legend shows the local probability of failure.}
\label{fig:Hazard}
\end{figure}

More insight into the local failure probability can be gained by means of a \emph{hazard map}. A hazard map highlights the areas being affected or vulnerable to a certain type of failure, providing visual information on the failure probability at each particular unit element \cite{MC16}. Thus, the local probability of failure (i.e., $P_f$ for a local $\sigma_w$) is shown in Fig. \ref{fig:Hazard} in logarithmic scale. The mesh can be easily constructed by reading the nodal coordinates and the element connectivity through \emph{Abaqus2Matlab}.

Statistical tools are indispensable to assess cleavage fracture as experimental data tends to be widely scattered; two identical specimens of the same material may have very different toughness values due to the random location of the fracture-triggering particle. \emph{Abaqus2Matlab} enables the usage of Matlab's in-built statistical capabilities to estimate all Weibull parameters without any prior assumptions. This novel iterative framework allows for more precise estimations of failure probabilities, a crucial aspect in risk quantification and operational decision making in engineering applications. 

\subsection{Cohesive zone modeling}

\begin{figure}[H]
\centering
\includegraphics[scale=0.6]{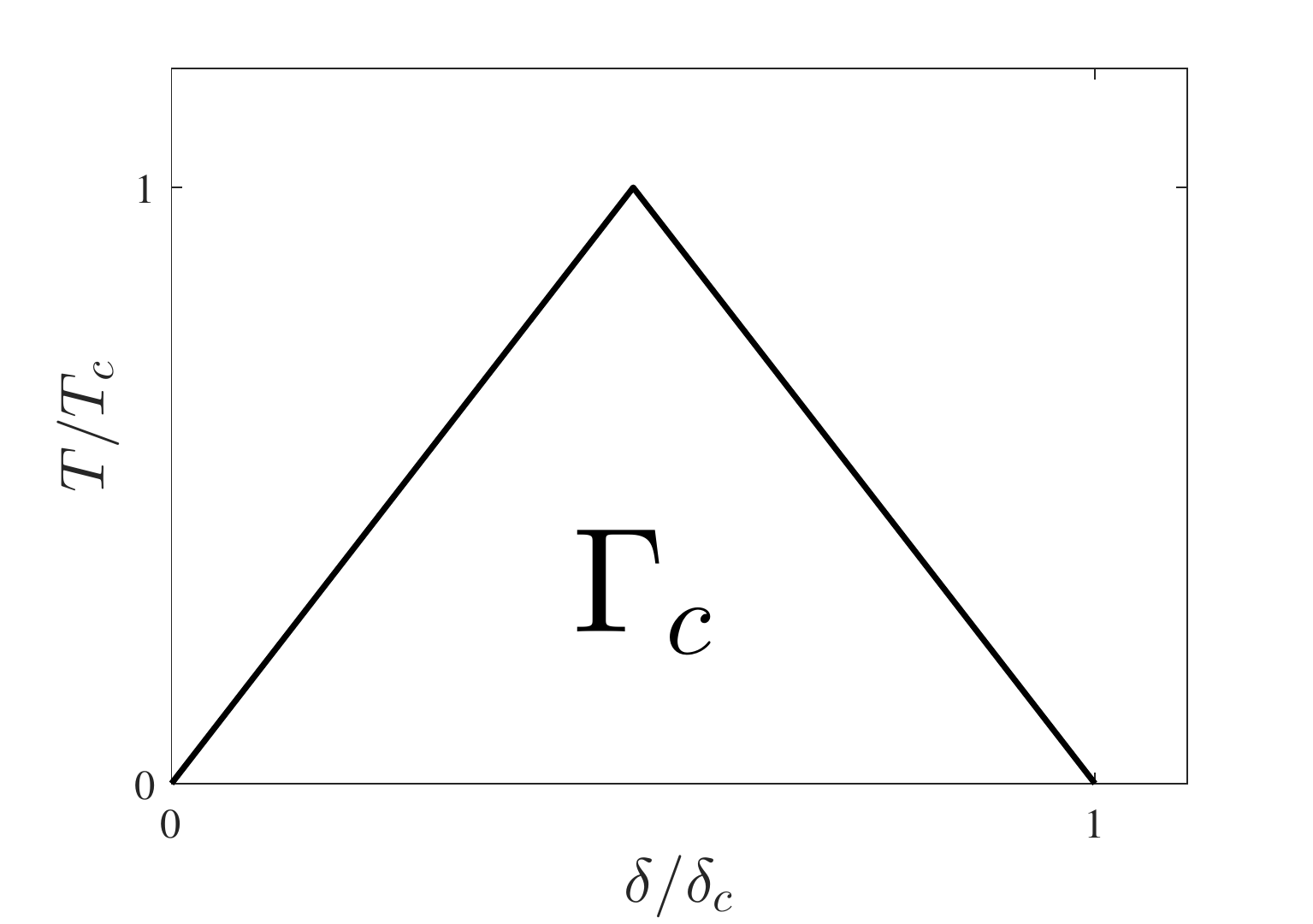}
\caption{Bi-linear traction separation law characterizing the cohesive zone model.}
\label{fig:CoheLaw}
\end{figure}

Classic fracture mechanics is an indispensable design tool that provides the basis for structural integrity assessment in engineering standards. The need to design components that exploit material performance to its maximum has however shifted scientific research from stationary cracks to crack propagation and damage. This is particularly true in ductile metals or composites, where a stable crack propagation stage precedes catastrophic failure. Among the many damage mechanics tools available, cohesive zone models are particularly attractive to characterize the reserve strength of the system once cracking has occurred, and to design accordingly \cite{C03}. The pivotal ingredient of cohesive zone modeling is the traction-separation law that governs material degradation and separation. As depicted in Fig. \ref{fig:CoheLaw}, for a given shape of the traction-separation curve, the cohesive response can be fully characterized through two parameters, the cohesive energy $\Gamma_c$ and the critical cohesive strength $T_c$. Thus, for the bi-linear law of Fig. \ref{fig:CoheLaw}, the cohesive energy can be expressed as a function of the critical separation $\delta_c$ and the critical cohesive strength $T_c$,
\begin{equation}
\Gamma_c=\frac{1}{2} T_c \delta_c
\end{equation}

The two parameters governing the cohesive response can be inferred from experiments. Generally, a trial and error procedure is followed, but such methodology is time consuming and error-prone. Here, a novel technique that builds on inverse analysis and neural network optimization is proposed to estimate the parameters governing the traction-separation law. \emph{Abaqus2Matlab} enables such an approach, not only by linking the advanced optimization tools available in Matlab with Abaqus damage modeling outcomes, but also by allowing to read and modify Abaqus input files accordingly. Thus, not only is \emph{Abaqus2Matlab} useful for post-processing purposes but it can be also used to optimize and pre-process through a \emph{two-way} interaction between Abaqus and Matlab. This is done by creating a Matlab function that reads the input file from Abaqus (*.inp) and, for this particular example, overwrites the line where the magnitude of $\delta_c$ and $T_c$ are defined. The function can be downloaded from \emph{Abaqus2Matlab}'s website and easily re-written to edit any other specific command.\\

\begin{figure}[H]
\centering
\includegraphics[scale=0.4]{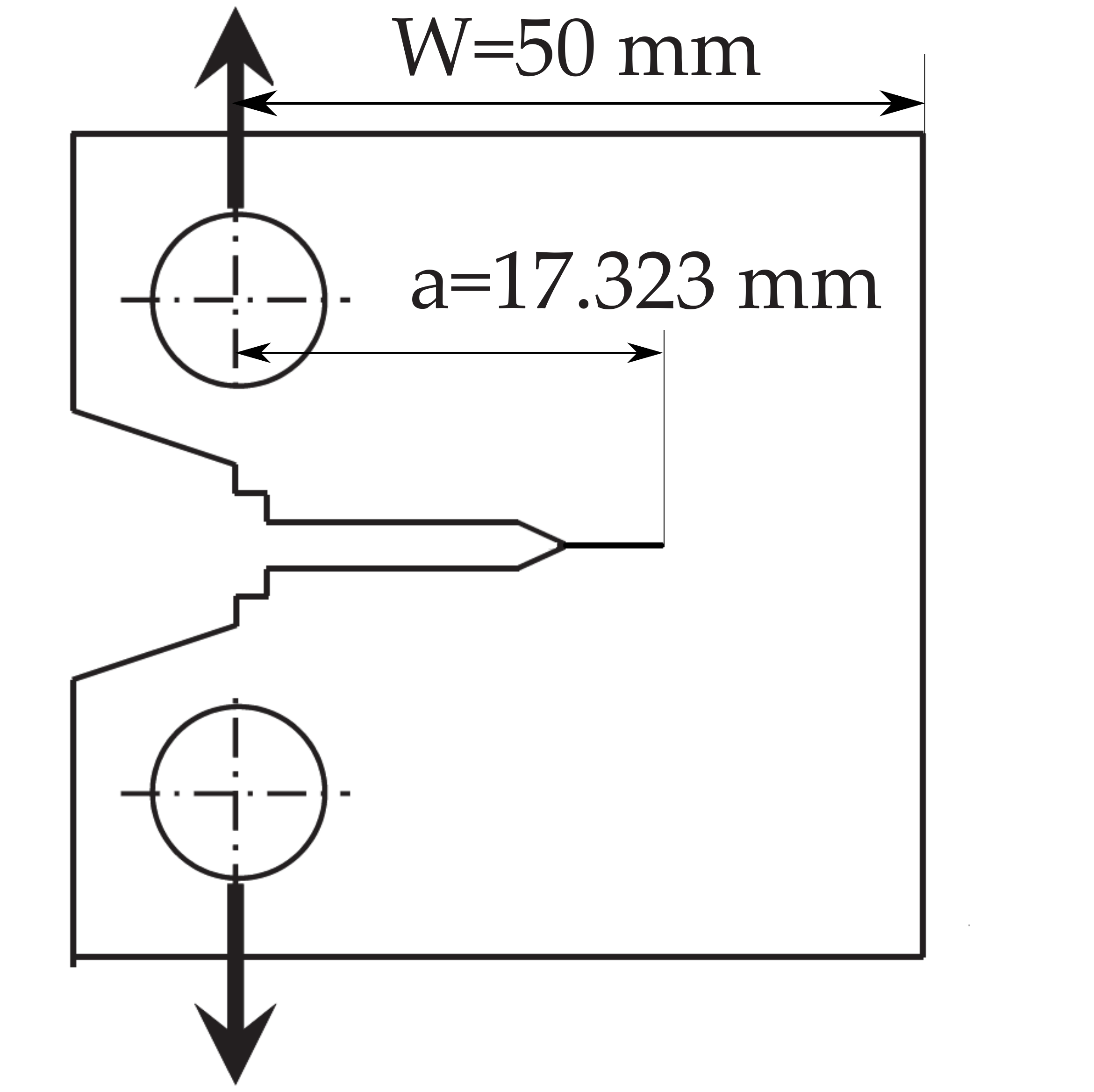}
\caption{Geometry and dimensions of the Al2024 Compact Tension specimen.}
\label{fig:CT}
\end{figure}

The material under consideration in the present study is Aluminum 2024. Both uniaxial and Compact Tension tests have been performed. The former lead to a Young's modulus of $E=85826$ MPa (Poisson's ratio $\nu=0.33$) while the plastic behavior can be fitted through a Hollomon's law $\sigma=k \varepsilon_p^n$ with $k=733$ and $n=0.157$. As depicted in Fig. \ref{fig:CT}, the specimen has a width of $W=50$ mm, a thickness of $B=20$ and a total crack length of $a=17.323$ mm, being the fatigue pre-crack equal to $a_0=7.323$ mm.

\begin{figure}[H]
\centering
\includegraphics[scale=0.8]{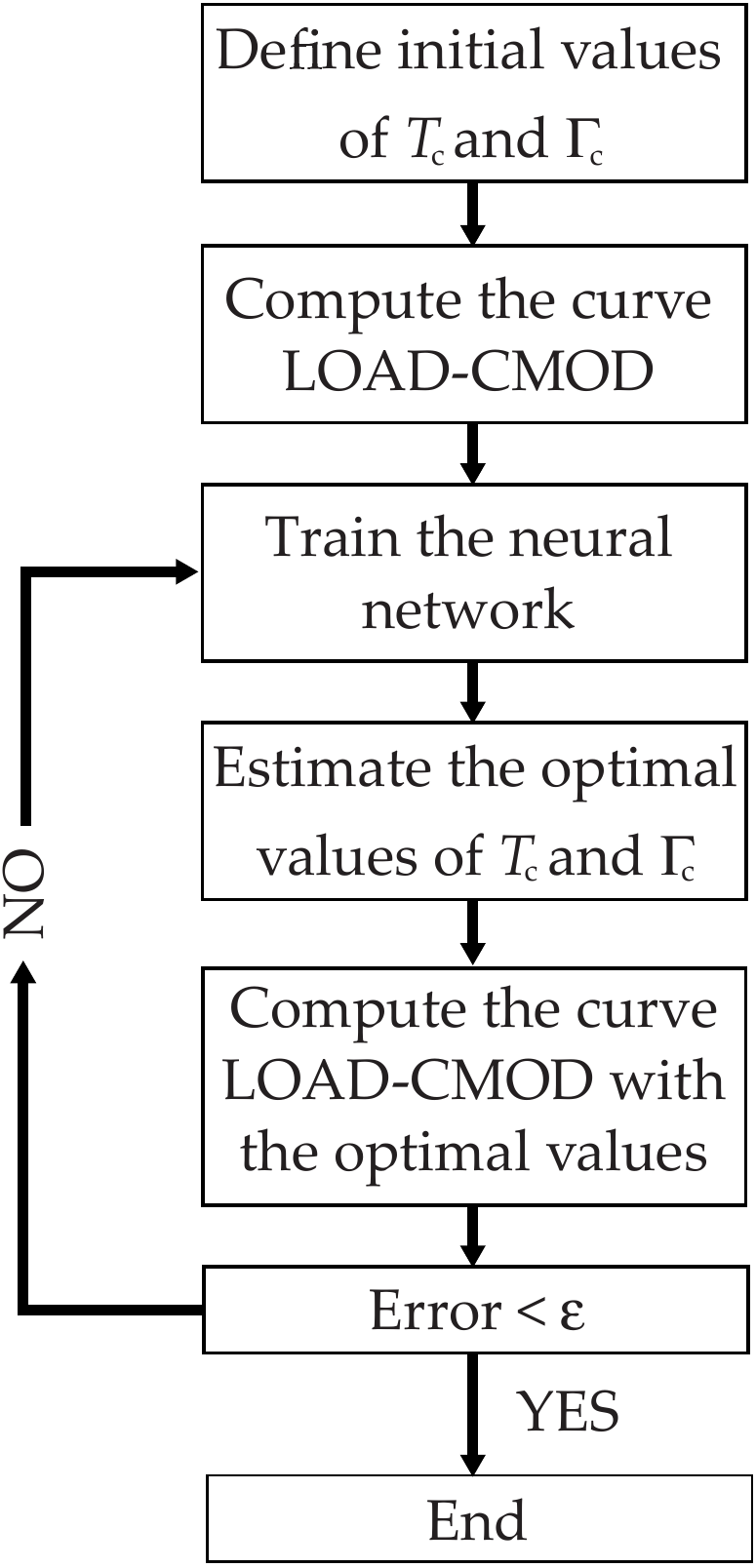}
\caption{Neural network optimization flowchart.}
\label{fig:OptAlgorithm}
\end{figure}

\begin{figure}[H]
        \raggedright
        \begin{subfigure}[h]{0.49\textwidth}
                \centering
                \includegraphics[scale=0.55]{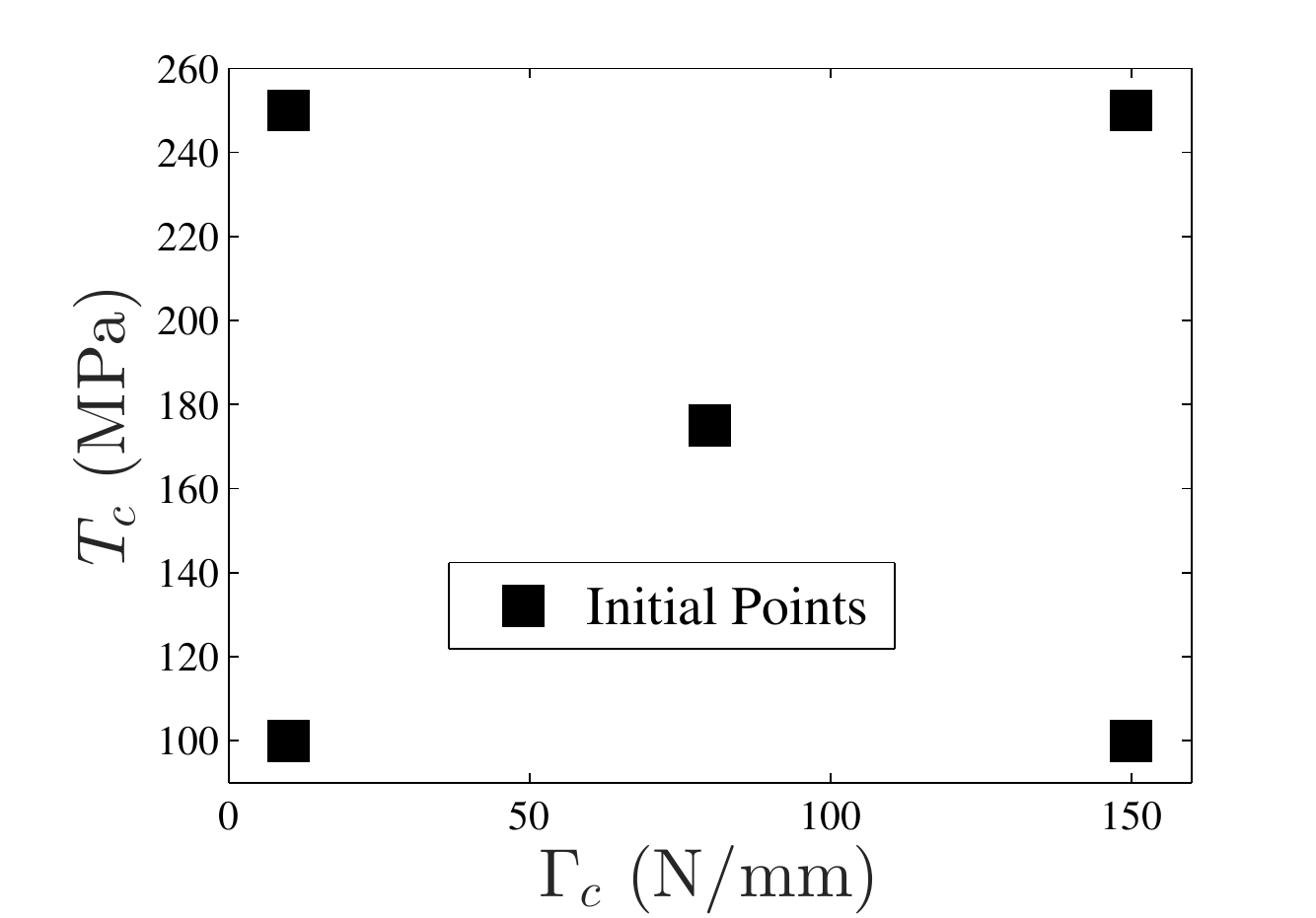}
                \caption{}
                \label{fig:InitialPoints}
        \end{subfigure}
        \begin{subfigure}[h]{0.49\textwidth}
                \centering
                \includegraphics[scale=0.55]{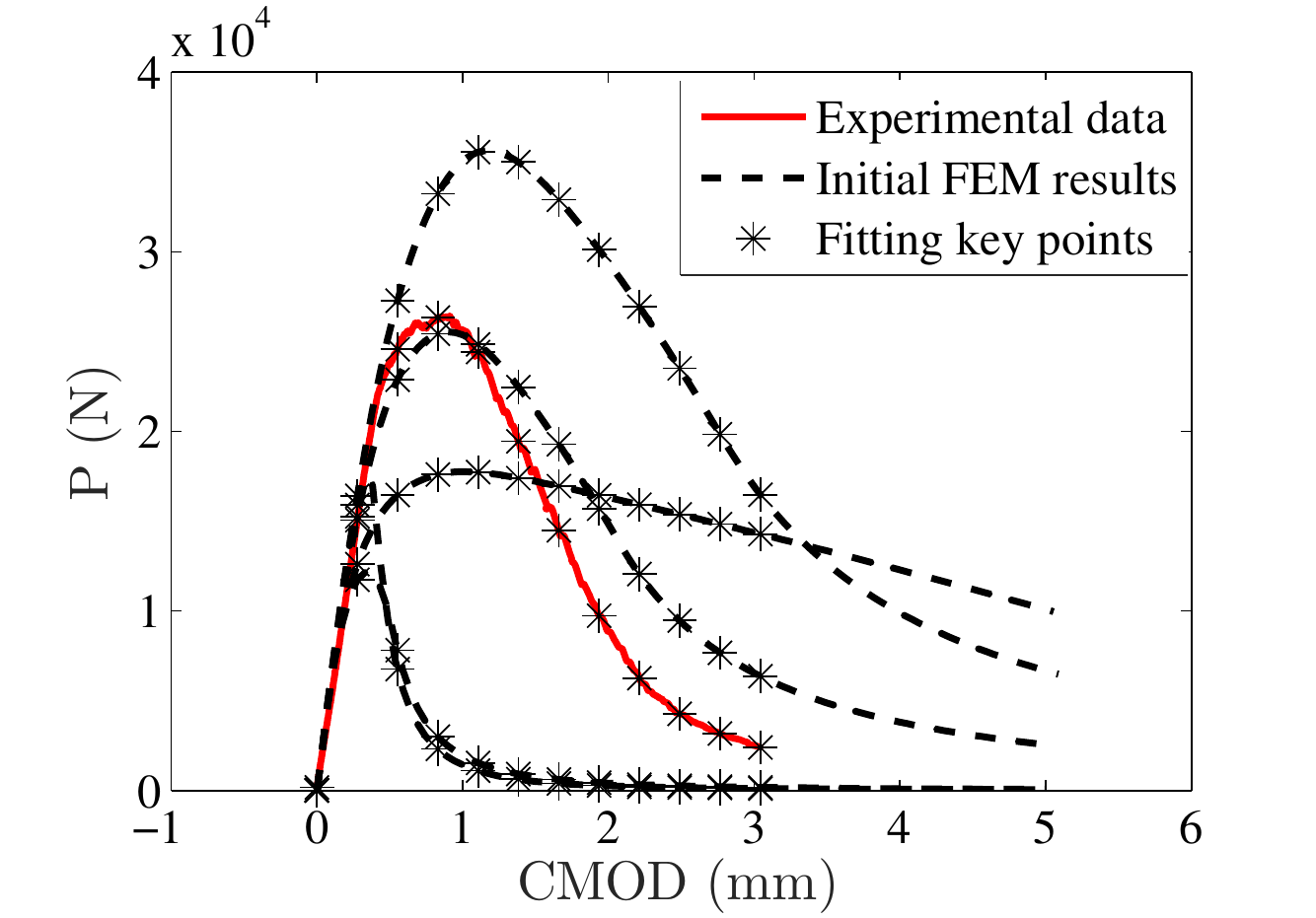}
                \caption{}
                \label{fig:LoadCMOD}
        \end{subfigure}
       
        \caption{Initial neural network training steps, (a) First set of $T_c$-$\Gamma_c$ values, and (b) corresponding load versus crack mouth opening displacement curves.}\label{fig:InitialCZM}
\end{figure}

The optimization procedure proposed correlates numerical results and experimental data of load versus crack mouth opening displacement (CMOD) by following the flowchart shown in Fig. \ref{fig:OptAlgorithm}. Thus, the first step involves assigning a set of initial values to $T_c$ and $\Gamma_c$. These initial values should be chosen so as to span a considerably wide range, ensuring that the optimal solution falls inside. The more numerous the merrier, as the performance of the neural network increases with the number of points. Nevertheless, only 5 pairs of $T_c$ vs $\Gamma_c$ points will be employed in this example to show the model capabilities even with a few initial values (see Fig. \ref{fig:InitialPoints}).

The finite element calculations are then performed, where Abaqus capabilities to model cohesive zone damage are employed and a very refined mesh of quadrilateral quadratic plane strain elements with reduced integration is adopted. The curve load versus CMOD is obtained in \emph{Abaqus2Matlab} by reading the nodal reaction forces and the displacement in particular sets (\mcode{Rec104} and \mcode{Rec101} functions). Computations are efficiently performed for each pair of $T_c$-$\Gamma_c$ values by taking advantage of \emph{Abaqus2Matlab} capabilities to read and modify Abaqus' input file. The results obtained in each case are shown in Fig. \ref{fig:LoadCMOD}; each curve is characterized by 12 equally distant points so as to correlate with the experimental data.

\begin{figure}[H]
\centering
\includegraphics[scale=0.9]{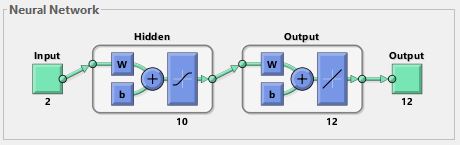}
\caption{Graphical summary of the characteristics of the Neural Network employed.}
\label{fig:NeuralNetwork}
\end{figure}

The next step involves training the neural network based on the input ($T_c$ and $\Gamma_c$ values) and output (load versus CMOD curves) information. The network is composed of 10 hidden layers and is trained by employing the Bayesian Regulation Method available in Matlab (see Fig. \ref{fig:NeuralNetwork}); full advantage of the \emph{Neural Net Fitting Matlab App} can be gained with \emph{Abaqus2Matlab}. In this example, 80\% of the models have been employed to train the network, 15\% of them have been used for validation purposes and the remaining 5\% serve to test the final solution obtained.

\begin{figure}[H]
\centering
\includegraphics[scale=0.8]{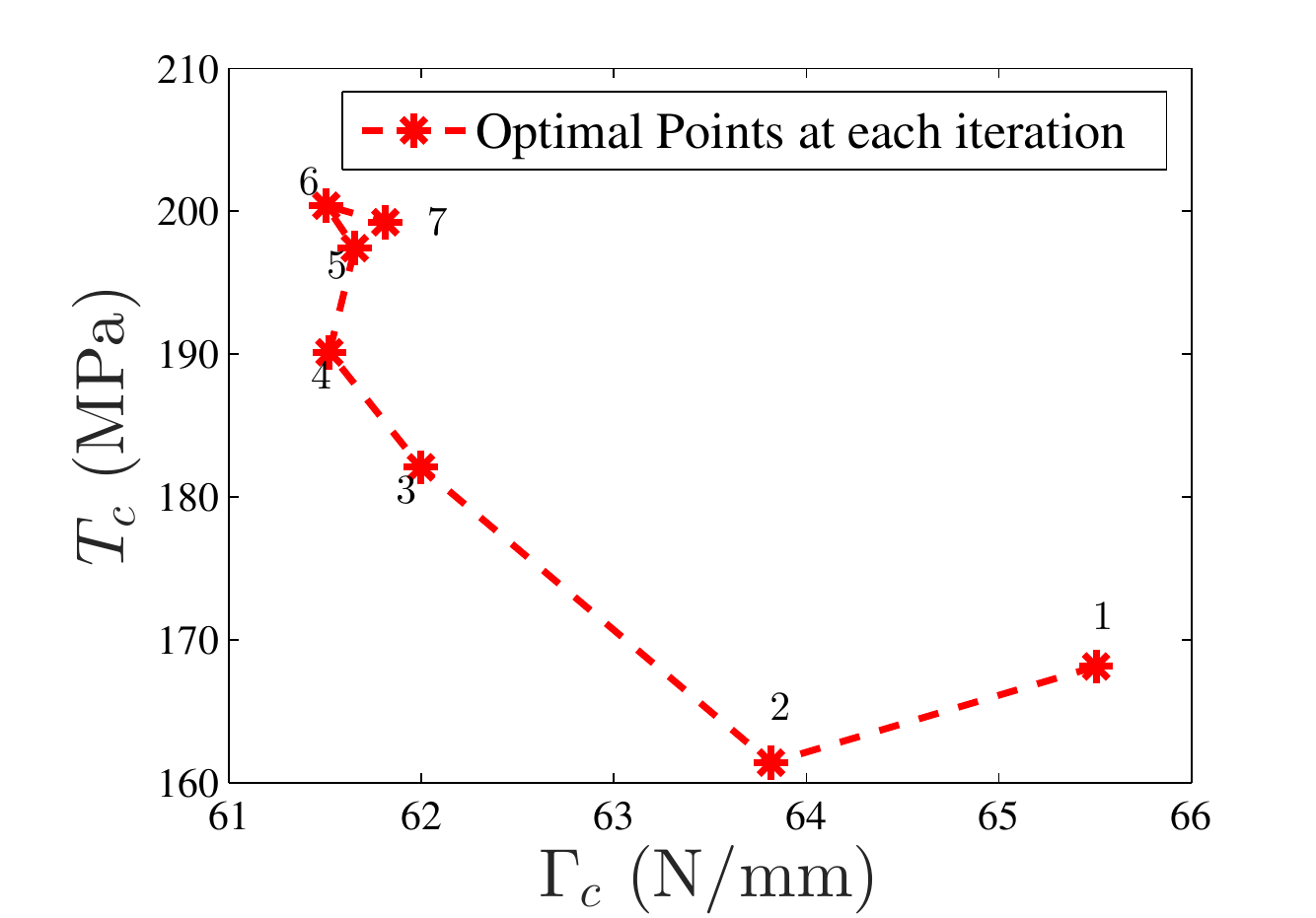}
\caption{Cohesive strength $T_c$ and fracture energy $\Gamma_c$ estimations at each iteration.}
\label{fig:OptimalPoints1}
\end{figure}

Once the neural network is fitted and tested, it is used to estimate - through least squares fitting - the optimal values of $T_c$ and $\Gamma_c$ by minimizing the differences between the load-CMOD curve obtained from the model and its experimental counterpart. To assess the quality of the neural network prediction, the optimized values of the cohesive strength and the cohesive fracture energy are provided as input to the finite element model. The outcome of this new finite element analysis is compared to the experimental data. If the norm of the differences between the curves is higher than a given tolerance, the neural network is trained again by adding new input and output information from the previous iteration. Fig. \ref{fig:OptimalPoints1} shows the optimal values of the strength and the cohesive energy obtained in each iteration.

\begin{figure}[H]
\centering
\includegraphics[scale=0.8]{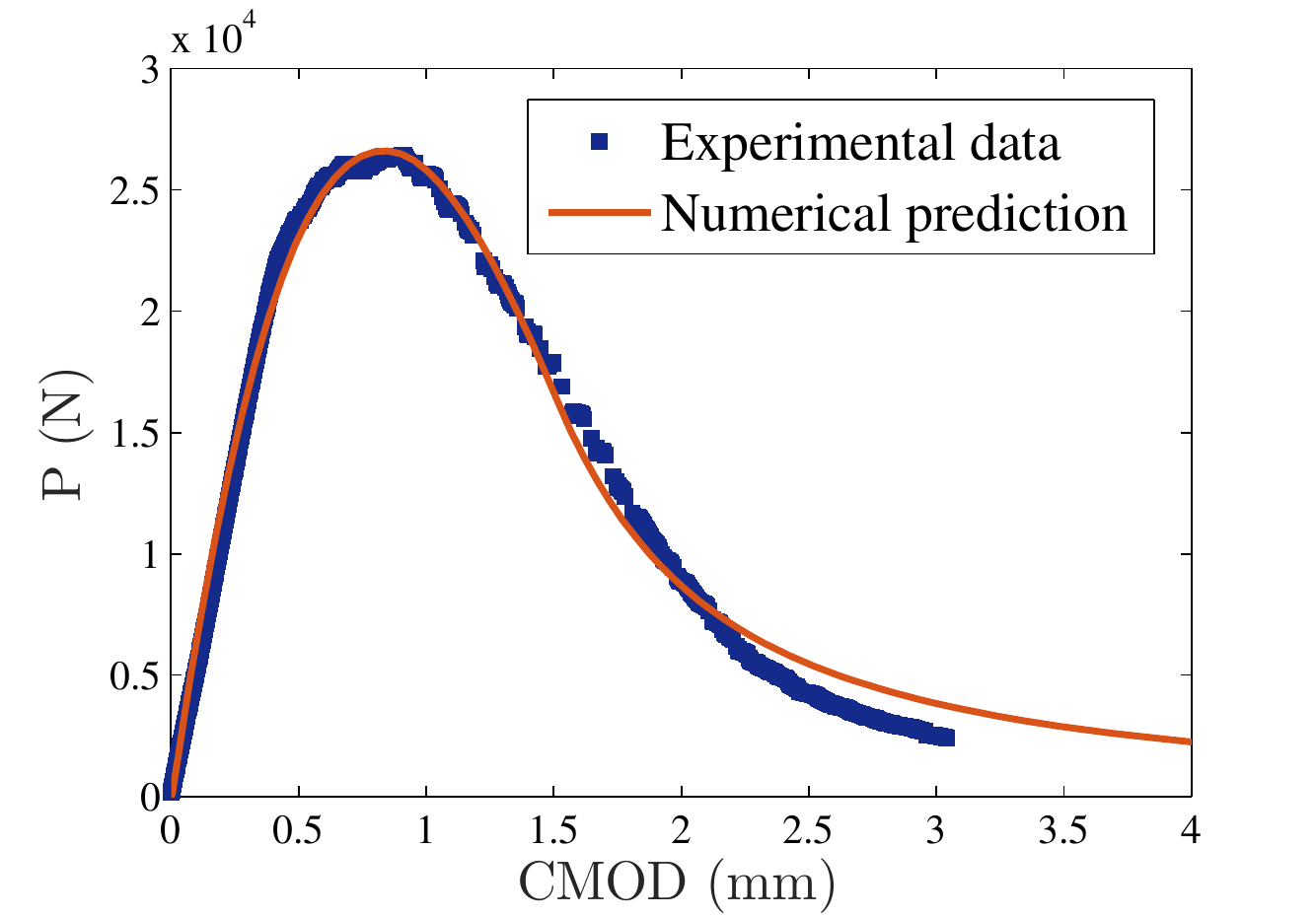}
\caption{Experimental and numerically optimized predictions of load versus crack mouth opening displacement in Al2024.}
\label{fig:LoadCMODfinal}
\end{figure}

In the present example convergence is achieved after 7 iterations and the final outcome is shown in Fig. \ref{fig:LoadCMODfinal}, together with the experimental result. As it can be seen in the figure, the optimal values ($T_c=199.2$ MPa and $\Gamma_c=61.81$ N/mm) lead to a very good quantitative agreement with the load versus CMOD curve obtained experimentally.

Hence, quantitative insight into the initiation and subsequent propagation of damage can be obtained through neural network optimization and a hybrid experimental-numerical strategy, in what is usually referred to as a top-down approach \cite{Ma16,Mb16}. Thus, \emph{Abaqus2Matlab} largely facilitates structural integrity assessment by taking advantage of advanced damage models available in Abaqus and modern optimization capabilities of Matlab. Moreover, its usage can be easily extended to a wide range of non-linear problems, where inverse analysis is an indispensable tool.\\

A detailed description, and the associated codes of the novel approaches employed in the two challenging engineering problems addressed, can be downloaded from \emph{Abaqus2Matlab} website.

\section{Conclusions}
\label{Sec:Concluding remarks}

A novel toolbox has been presented to couple Abaqus and Matlab. Its development is motivated by the need of an open source package that provides a non-intrusive link between the highly-developed finite element capabilities of Abaqus and the comprehensive analysis tools of Matlab. The software, conveniently titled \emph{Abaqus2Matlab}, unfolds an extensive range of modeling possibilities. Its capabilities are particularly attractive from the post-processing perspective, enabling to complement advanced finite element simulations with the numerous graphical and mathematical analysis options of Matlab and its toolboxes.\\

Two practical cases with important implications in structural integrity assessment are investigated to illustrate the potential of \emph{Abaqus2Matlab}. First, cleavage fracture is examined by means of a three-parameter Weibull approach. A novel statistical framework is proposed to estimate the modulus, the scaling and the threshold parameters through \emph{Abaqus2Matlab} without any preceding assumptions. The software is also employed to model crack propagation in Al2024 by extracting the cohesive parameters from the experimental data through inverse analysis. \emph{Abaqus2Matlab} plays a fundamental role by enabling model manipulation and genetic algorithm optimization. The general structure of the code facilitates its application to numerous engineering problems with minimum coding effort. Diverse examples (including the ones described here), comprehensive documentation and the source code can be downloaded from www.abaqus2matlab.com. 

\section{Acknowledgments}
\label{Acknowledge of funding}

T.E. Garc\'{\i}a (University of Oviedo) is acknowledged for his work on the experimental part of the manuscript. E. Mart\'{\i}nez-Pa\~neda acknowledges financial support from the People Programme (Marie Curie Actions) of the European Union's Seventh Framework Programme (FP7/2007-2013) under REA grant agreement n$^{\circ}$ 609405 (COFUNDPostdocDTU).



\end{document}